\documentstyle[preprint,revtex]{aps}
\begin{document}
\voffset= -0.5 in
\vsize = 24.0truecm
%% BEGIN TITLE %%%%%%%%%%%%%%%%%%%%%%%%%%%%%%%%%%%%%%%%%%%%%%%%%%
\begin{center}
{\bf NUCLEAR BREATHING MODE IN THE RELATIVISTIC MEAN-FIELD THEORY} \\
\vspace{0.3cm}

M.V.Stoitsov$^{1,*}$,
M.L. Cescato$^{1,3}$, P. Ring$^1$ and M.M. Sharma$^2$

\vspace{0.3cm}
$^1$Physik Department, Technische Universit\"at M\"unchen, D-85748 Garching,
Germany.\\
$^2$Max Planck Institut f\"ur Astrophysik, Karl-Schwarzschild-Strasse 1,
D-85740 Garching, Germany.\\
$^3$Departamento de F\'\i sica - CCEN, Universidade Federal da Para\'\i ba \\
C.P. 5008, 58051-970 Jo\~ao Pessoa - PB, Brasil.\\

\end{center}
\vspace{1cm}

%% BEGIN ABSTRACT %%%%%%%%%%%%%%%%%%%%%%%%%%%%%%%%%%%%%%%%%%%%%%%%%%

\begin{center}
{\bf Abstract}
\end{center}
\baselineskip=18pt
The breathing-mode giant monopole resonance is studied within
the framework of the relativistic mean-field (RMF) theory. Using a broad
range of parameter sets, a systematic analysis of constrained
incompressibility and excitation energy of isoscalar monopole states
in finite nuclei is performed. A comparison is made
with the incompressibility derived from the semi-infinite
nuclear matter and with constraint nonrelativistic Skyrme Hartree-Fock
calculations. Investigating the dependence of the breathing-mode
energy on the nuclear matter incompressibility, it is shown
that dynamical properties of surface respond differently in the
RMF theory than in the Skyrme approach.

\newpage
\baselineskip=17pt
% INTRODUCTION

The breathing-mode giant monopole resonace (GMR) has been a matter
of contention in the recent past\cite{sha90}. The energy of the GMR has
been considered to be a source of information on the nuclear matter
compressibility. Empirically, data
on the GMR have been measured with considerable precision\cite{sha88}.
The detailed analysis of the GMR leading to the incompressibility of
nuclear matter is as yet not settled and is still under investigation.
Theoretically, the incompressibility is understood to
have been obtained\cite{bla80} using the density-dependent
Skyrme interactions. The deductions base themselves upon interpolation
between various Skyrme forces for the GMR energies obtained from HF+RPA
calculations. There has been, however, no unambiguous reproduction of
the empirical GMR energies of medium-heavy nuclei using this approach.
The breathing-mode energies depend not only upon the bulk
incompressibility, but are also sensitive to the surface incompressibility.
The relationship between the bulk and the surface incompressibility has been
enunciated clearly for the Skyrme approach in Ref. \cite{tre81}.

The relativistic mean-field (RMF) theory has achieved a considerable
success\cite{snr93} in describing the ground-state properties of
nuclei at and far away from the stability line.
The dynamical aspects have, however, remained largely unexplored.
The breathing-mode energies and incompressibilities were obtained
within the RMF theory using the linear Walecka model\cite{mar89,malf91}.
The relationship of the GMR energies to the incompressibility of
nuclear matter is not yet clear for the RMF theory.
In the non-relativistic Skyrme approach, on the other hand,
the GMR energies are
connected to the incompressibility straightforwardly. In this letter,
we examine the dependence of the breathing-mode energies in finite
nuclei on the incompressibility of nuclear matter in the RMF theory.

%% BEGIN STATIC PART(THEORY)%%%%%%%%%%%%%%%%%%%%%%%%%%%%%%%%%%%%%%%%%

  We start from a relativistic Lagrangian\cite{ser86} which treates
  nucleons as Dirac spinors $\psi$ interacting by the exchange of several
  mesons: scalar $\sigma$-mesons that produce a strong attraction, isoscalar
  vector $\omega$-mesons that cause a strong  repulsion and isovector
  $\rho$-mesons required to describe the isospin asymmetry.
  Photons provide the necessary electromagnetic interaction.
  The model Lagrangian density is:

\begin{equation}
\begin{array}{ll}
{\cal L}& =  \bar\psi \{i\gamma_\mu\partial^\mu-M\}\psi
   +{1\over 2}\partial^\mu\sigma\partial_\mu\sigma
   -U(\sigma)-g_\sigma\bar\psi\sigma\psi \\
  & -{1\over 4}\Omega^{\mu\nu}\Omega_{\mu\nu}
   +{1\over 2}m^2_\omega \omega_\mu\omega^\mu
   -g_\omega\bar\psi\gamma_\mu\omega^\mu\psi  \\
  & -{1\over 4}\vec {R}^{\mu\nu}\vec {R}_{\mu\nu}
   +{1\over 2}m^2_\rho \vec {\rho}_\mu\vec {\rho}^\mu
   -g_\rho\bar\psi\gamma^\mu\vec{\tau}\psi\vec{\rho}_\mu  \\
  & -{1\over 4}\vec {F}^{\mu\nu}\vec {F}_{\mu\nu}
   -e\bar\psi\gamma^\mu{{(1-\tau_3)} \over 2} \psi A_\mu,
\end{array}
\end{equation}
where $U(\sigma)$ is the non-linear potential with the cubic and
quartic terms\cite{bog77}:

\begin{equation}
U(\sigma)={1\over 2}m_\sigma\sigma^2+
{1\over 3}g_2\sigma^3+{1\over 4}g_3\sigma^4.
\end{equation}
M, $m_\sigma$, $m_\omega$ and $m_\rho$ are the nucleon, the $\sigma$-, the
$\omega$-, and the $\rho$-meson masses, respectively, and $g_\sigma$,
$g_\omega$, $g_\rho$ and $e^2$/$4\pi$=1/137 are the coupling
constants for the $\sigma$-,  the $\omega$-, the $\rho$-mesons and for the
photon. The field tensors for the vector mesons are
$\Omega^{\mu\nu}=\partial^\mu\omega_\nu-\partial^\nu\omega_\mu$,
$\vec {R}^{\mu\nu}=\partial^\mu\vec{\rho}_\nu-\partial^\nu\vec{\rho}_\mu
                          - g_\rho(\vec {\rho}^\mu \times \vec{\rho}_\nu)$
and  for the electromagnetic field
${F}^{\mu\nu}=\partial^\mu{A}_\nu-\partial^\nu{A}_\mu$. The
variational principle leads to the stationary Dirac equation with
the single-particle energies as eigenvalues,

\begin{equation}
\hat h_D \psi_i(x) = \varepsilon_i \psi_i(x),
\end{equation}
where

\begin{equation}
\begin{array}{ll}
\hat h_D  = -i\vec\alpha.\bigtriangledown
            + \beta (M + g_\sigma \sigma(r))
+g_\omega \omega^0(r) +g_\rho \tau_3 \rho^0(r)
            + e{{(1-\tau_3)} \over 2}A^0(r).
\end{array}
\end{equation}
Solving these equations self-consistently one obtains the nuclear ground
state in terms of the solution ${\psi_i}$.

%% BEGIN DYNAMIC PART(THEORY)%%%%%%%%%%%%%%%%%%%%%%%%%%%%%%%%%%%%%%%%%
In order to obtain the isoscalar monopole states in nuclei we perform
constrained RMF calculations solving the Dirac equation

\begin{equation}
\left(\hat h_D - \lambda r^2\right) \psi_i(x,\lambda)
= \varepsilon_i \psi_i(x,\lambda),
\end{equation}
for different values of the Lagrange multiplier $\lambda$ which keeps the
nuclear $rms$ radius fixed at its particular value
\begin{equation}
R = \left\{{1\over A} \int r^2 \rho_\lambda(r) d^3r \right\}^{1/2},
\end{equation}
where
\begin{equation}
 \rho_\lambda(r)\equiv \rho_v(r,\lambda) =
 \displaystyle{\sum_{i=1}^A}\;\psi_i^\dagger(x,\lambda ) \psi_i(x,\lambda)
\end{equation}
is the local baryon density determined by the solution ${\psi_i(x,\lambda)}$.
The total energy of the constrained system
\begin{equation}
E_{RMF}(\lambda ) = E_{RMF}[\psi_i(x,\lambda )] ,
\end{equation}
is a function of $\lambda$ (or the $rms$ radius R) which has a minimum,
the ground state energy $E_{RMF}^0=E_{RMF}(0)$, at $\lambda =0$ corresponding
to the ground-state $rms$ radius $R_0$.

 This behaviour of the constrained energy (8) as a function of $\lambda$
 (or R) allows us to examine the isoscalar monopole motion of a nucleus as
 harmonic (breathing) vibrations changing the $rms$ radius R around its
 ground-state value $R_0$. Considering $s=(R/R_0-1)$ as a dynamical collective
 variable and expanding (8) around the ground-state point s=0 (or $\lambda=0$)
 we obtain in the harmonic approximation,
\begin{equation}
E_{RMF}(\lambda ) = E_{RMF}^0 + {1\over 2} A K_C(A) s^2,
\end{equation}
where
\begin{equation}
K_C(A) =
 A^{-1} \left( R^2 {{d^2E_{RMF}(\lambda )}\over {dR^2}} \right)_{\lambda=0},
\end{equation}
is the constrained incompressibility\cite{jenn80} of the finite nucleus.
The second term in eq.(9) represents the restoring force of the monopole
vibration. In order to obtain its associated inertial parameter we apply the
method\cite{mar89} of making a local Lorentz boost on the constrained spinors
 $\psi_i(x,\lambda )$:
\begin{equation}
\psi_i(x,t) = {1\over {\sqrt \gamma }} \hat S(\vec v) \psi_i(x,\lambda),
\end{equation}
where $ \hat S(\vec v)$ is the Hermitian local Lorentz boost operator
\begin{equation}
 \hat S(\vec v)= \cosh\left( {{\phi}\over 2} \right)
  + \vec{\alpha}.{{\vec v}\over {|\vec v|}} \sinh\left( {\phi \over 2}\right),
\end{equation}
with $\phi=tanh^{-1}(|\vec v|)$ and $\gamma=\sqrt{1-{\vec v}^2}$. The
velocity field $\vec v=\vec v(\vec r,t)$ is then obtained by the continuity
equation
\begin{equation}
{{\partial}\over {\partial t}}\; \displaystyle{\sum_{i=1}^A}
              \psi_i^\dagger(x,\lambda ) \psi_i(x,\lambda ) +
 \bigtriangledown .\displaystyle{\sum_{i=1}^A}
              \psi_i^\dagger(x,\lambda )\vec \alpha \psi_i(x,\lambda ) =0,
\end{equation}
which, using eqs.(7) and (11), transforms into the form
\begin{equation}
 \dot s \; {{\partial \rho (r,\lambda )}\over {\partial s}}
    +\bigtriangledown .\left( \rho (r,\lambda ) \vec v(r,t) \right) = 0.
\end{equation}
Up to first order in $\dot s$ the velocity field is determined by eq.(14) into
the form  $\vec v = - \dot s u(r) \vec r/|\vec r|$ where
\begin{equation}
u(r) = \left\{
   {\displaystyle{\int_0^r {{d\rho(r',\lambda )}\over {ds}} r'^2 dr'}} \over
         {\rho (r,\lambda) r^2 } \right\} .
\end{equation}
The inertial parameter of the monopole vibration is then obtained
as\cite{mar89}
\begin{equation}
B_{rel}(A) = A^{-1} \int u^2(r) \, {\cal H}_{RMF}(r)\, d^3r,
\end{equation}
where $u(r)$ is the velocity function (15) at $\lambda=0$ and
 $ {\cal H}_{RMF}(r)$ is the Hamiltonian density.

%% BEGIN RESULTS PART(CALCULATIONS)%%%%%%%%%%%%%%%%%%%%%%%%%%%%%%%%%%%

We have obtained $K_C(A)$ from eq.(10), $B_{rel}(A)$ from eq.(16)
and the frequency of  the  isoscalar monopole vibration as

\begin{equation}
 \omega_C = \sqrt{{K_C(A)\over B_{rel}(A)}}
\end{equation}
for a number of spherical nuclei. Various parameter sets such as
NL1\cite{rei89}, NL-SH\cite{snr93}, NL2\cite{lee86}, HS\cite{hor81}
and L1\cite{lee86} with values of the nuclear
matter incompressibility $K_{NM}=\,$211.7, 354.95, 399.2, 545 and
626.3 MeV, respectively, have been employed in the calculations.
The last two sets, HS and L1, correspond to linear
models without the self-coupling of the $\sigma$-field.
In addition, the set L1
excludes the contribution from the $\rho$-field.
Among the sets NL1, NL-SH and
NL2 which correspond to the non-linear model, only the set NL2 has a positive
coupling constant $g_3$, eq(2). Whereas the set NL1 reproduces the
ground-state properties of nuclei only close to the stability line due to
the very large asymmetry energy, the set NL-SH describes also nuclei
very far away from the stability line\cite{sha94b}.

Results from the present constrained RMF calculations are shown in
Figs. 1 and 2, where the collective mass $B_{rel}(A)$, the constrained
incompressibility $K_C(A)$ and the associated excitation energies
$\omega_C$  for a few nuclei have been displayed. First, we
consider $K_C(A)$ in Fig. 1 (a).
The incompressibility of nuclei shows a strong dependence on the nuclear
matter incompressibility $K_{NM}$, with a few exception for light nuclei.
For the linear force HS, $K_C$ shows a slight dip from the increasing trend
for $^{208}$Pb and $^{90}$Zr, whereas for light nuclei $^{40}$Ca and $^{16}$O
the HS values are even smaller than the NL2 values. The dependence of
the imcompressibility $K(A)$ of finite nuclei on $K_{NM}$ obtained from
non-relativistic Skyrme calculations is different: it
increases monotonically with $K_{NM}$\cite{bla80}.
This difference can be understood from the difference in the
behaviour of the surface incompressibility in the two methods. In the
Skyrme approach, the surface incompressibility has been shown to be
$K_S \sim -K_{NM}$ for standard Skyrme forces\cite{tre81}.
However, this does not seem to be the case for the RMF theory as
shown by the dip at the HS values. Thus, the surface incompressibility
is not necessarily a straight
function of the nuclear matter incompressibilty in the RMF theory.

Since the GMR energy depends strongly upon the inertial mass parameter,
we next examine the mass parameter $B_{rel}(A)$ in Fig. 1 (b). The
collective mass obtained in Eq. 16 from the velocity field shows an
interesting dependence on the mass of the nucleus considered.
For the heavy nucleus
$^{208}$Pb, the collective mass is about 0.9 and for $^{90}$Zr it is
0.5 MeV$^{-1}$. For lighter nuclei it decreases to about 0.3 MeV$^{-1}$,
as shown in Table 1. It does not,
however, depend much upon the parameter set used and therefore
shows only little sensitivity to $K_{NM}$. This implies that the
$K_{NM}$ dependence of the energy is then predominantly due to the
$K_{NM}$ dependence of the incompressibility  $K_C(A)$.

We now compare in Table 1 the results from two RMF parameter sets,
NL1 and NL-SH. The constrained incompressibilities $K_C$ for
NL-SH are higher than for NL1 as discussed above in Fig. 1(a).
It is interesting to compare the relativistic collective mass
$B_{rel}(A)$ with the expression usually applied in
the non-relativistic sum-rule approach,
\begin{equation}
 B_{sr}(A)= M R_0^2,
\end{equation}
where  the ground-state $rms$ radius $R_0$ is used. The ratio
$B_{sr}(A)/B_{rel}(A)$ is shown in Table 1 where the energy
$\omega_C$, eq.(17) and its approximation
\begin{equation}
 \omega_{sr} = \sqrt{{K_C(A)\over B_{sr}(A)}}
\end{equation}
have also been compared for the sets NL1 and NL-SH. The observation
that the collective mass $B_{rel}(A)$ decreases significantly (up
to 50 \%) for light nuclei, provides a hint for the influence of the
surface in the value $B_{rel}(A)$. For heavy nuclei the approximate value
$B_{sr}(A)$ is rather close to $B_{rel}(A)$.
Consequently, the monopole energies $\omega_C$ and
$\omega_{sr}$ differ by about 1 MeV only for heavy nuclei and
up to about 5 MeV for the lighter ones.

The energy $\omega_{sr}$ corresponds to the monopole excitation energy $E_1$
usually obtained from nonrelativistic Skyrme forces within the sum rule
approach\cite{boh79} and in nonrelativistic constrained Hartree-Fock
calculations.
In Table 1 energies $\omega_C$ and $\omega_{sr}$ have been compared with
such nonrelativistic constrained HF results $\omega_{Sky}$ obtained from
Eq. (19) with the
Skyrme-type forces SkM and SIII having about the same nuclear matter
incompressibility $K_{NM}$ as the sets NL1 and NL-SH, respectively.
 From Table 1 it can be seen that the Skyrme results\cite{tre81}
$\omega_{Sky}$ are actually close to the energies $\omega_{sr}$
and differ significantly from the values of $\omega_C$.
This difference in the RMF constrained energy $\omega_C$
from the Skyrme constrained energy $\omega_{Sky}$ is small
for heavy nuclei. It, however, increases for lighter nuclei, where the
RMF shows smaller values. The difference is particularly significant
for the higher compressibility forces (NL-SH and SIII). It arises naturally
from the lower values of the collective mass $B_{rel}(A)$ for light nuclei in
the RMF theory as discussed above. Thus, the masses $B_{rel}$ and $B_{sr}$
have different dependences on surface.

Fig. 2 shows the constrained breathing-mode energy $\omega_C$ for
different nuclei. For heavier nuclei $^{208}$Pb and $^{90}$Zr, $\omega_C$
shows a behaviour similar to that shown by $K_C$ in Fig. 1 (a),
as the collective mass does not show much change with incompressibility
(Fig. 1.b). The stagnation in energy at NL2 and HS is reminiscent of
the effect of surface compression as in Fig. 1 (a),
thus reflecting the role played by the surface in the RMF
theory. In lighter nuclei this effect is more transparent.
For $^{40}$Ca and $^{16}$O, $\omega_C$ is not related to $K_{NM}$ in
a simple way due to the combined effect of $K_C$ and the collective
mass $B_{rel}$, where there is even a decrease in $\omega_C$ with $K_{NM}$.
Employing schematic parameter sets, it was also
shown\cite{snr94} earlier that within the relativistic quantum
hadrodynamics the surface compression responds differently than in
the Skyrme ansatz.

A comparison of the empirical values of the GMR energies with $\omega_C$
is worthwhile. The empirical values for $^{208}$Pb and $^{90}$Zr
are 13.9$\pm$0.3 and 16.4$\pm$0.4 MeV, respectively.
For lighter nuclei the values are
very uncertain. Systematics of the values for $^{208}$Pb show that
the empirical value is encompassed by the constrained calculations
curve from $K_{NM}$ = 300 - 400 MeV. The $^{90}$Zr value is, however,
overestimated by the corresponding curve by 2-3 MeV.

We have also carried out calculations of the incompressibility
for finite nuclei by scaling the density using semi-infinite
nuclear matter. The incompressibility $K_A$ for a finite nucleus
can be written as:
\begin{equation}
K_A = K_{NM} + K_S A^{-1/3} + K_{\Sigma} \Biggl({N-Z \over A}\Biggr)^2 +
 {3\over5} {e^2\over r_0} \Biggl(1 -27 {\rho_0^2e'''\over K_{NM}}\Biggr).
\end{equation}
where major contribution to $K_A$ arises from the volume ($K_{NM}$)
and the surface terms ($K_S$). The surface incompressibility $K_S$ has
been obtained by calculating the second derivative of the surface
tension ($\sigma$) with respect to the density for each change in the
scaled density as given by\cite{bla80}
\begin{equation}
K_S~=~4\pi r_0^2\bigl\{22\sigma(\rho_0)~+~
9\rho_0^2\sigma''(\rho_0)~+~
{{54\,\sigma(\rho_0)}\over{K_\infty}}\rho^3_0
e'''_{\infty}(\rho_0)\bigr\},
\end{equation}
The details of the procedure to perform scaling of the density
have been discussed in ref\cite{sto91}.
The last two terms in eq. (20) contribute very little. We add
these terms for completeness, however. The asymmetry coefficient $K_\Sigma$
has been taken at $-$300 MeV from the empirical determination\cite{sha89}
and is a reasonable value. The third derivative of the EOS, $\rho_0^2e'''$,
for each force is known from the nuclear matter calculations.
We have performed the calculations with the parameter sets
used above.

Fig. 3 shows the 'scaling' incompressibility $K_A$ for various
RMF forces obtained from Eq. (20).
The general trend of the scaling incompressibility with $K_{NM}$ is
about the same as in Fig. 1(a), showing a dip for the force HS.
In general, the $K_A$ values are about 20\% larger than the
constrained values $K_C$. This is consistent with the known relationship
of the two types of the incompressibilities for nuclear matter,
whereby it was shown\cite{jenn80} that $K_A (NM) = \frac{7}{10} K_C (NM)$.
Fig. 3 brings out again the importance of the role of the
surface in the dynamical calculations within the RMF theory.
The surface incompressibility $K_S$ for the forces NL1 and
NL-SH are $-333.1$ and $-610.1$ MeV respectively. Thus, in both
the cases the ratio of the surface to the bulk incompressibility
is obtained as 1.58 and 1.72 respectively. These values differ
considerably from the ratio of 1 in the Skyrme ansatz. The values
of the bulk and surface incompressibilities for NL-SH are closer
to the empirical values from Ref. \cite{sha88}.

In conclusion, it has been shown that within the RMF theory,
the incompressibility of a finite nucleus, $K(A)$, does not
depend on the nuclear matter incompressibility in a simple
way. This dependence is strongly related to the properties of the
surface subjected to compression. This is different
from the behaviour of the surface in the Skyrme ansatz. The collective
mass for the breathing mode vibration in the RMF theory
also shows a different behavior from light to heavy nuclei,
thus affecting the frequency of the isoscalar breathing mode.
\vspace{0.3cm}

This work is supported in part by the Bundesminiterium f\"ur Forschung
und Technologie, Germany under the project 06 MT 733.
M.V.S. is supported by the Contract $\Phi-32$
with the Bulgarian National Science Foundation and by the
EEC program {\it Go West} under the Contract ERB-CHBI-CT93-0651.
M.L.C. would like to acknowledge support from CNPq Brazil.

\newpage
\par\noindent
$^*$Permanent Address: Institute of Nuclear Research and Nuclear
Energy, Bulgarian Academy of Sciences, Blvd. Tzarigradsko Chossee 72,
Sofia 1784, Bulgaria.

%% BEGIN (TABLE ONE)%%%%%%%%%%%%%%%%%%%%%%%%%%%%%%%%%%%%%%%%%%
\newpage
\hoffset=-3.3 truecm
\begin{table}
\caption{ The constrained incompressibility $K_C(A)$\, in MeV,
the mass parameter $B_{rel}(A)$ in MeV$^{-1}$, the ratio
$B_{sr}(A)/B_{rel}(A)$ and the associated  monopole
frequencies  $\omega_C$  and $\omega_{sr}$, (both in MeV) calculated with the
sets NL1 and NL-SH. Comparison is made with constrained Skyrme results
$\omega_{Sky}$ obtained within the sum-rule approach\cite{boh79} with the
Skyrme forces SkM and SIII.}
\hsize=18.9truecm
\begin{tabular}{ccccccccccccccc}
 &\,\,&\multicolumn{6}{c}{{NL1:}  {$K_{NM}=211.7$\,MeV}}
                                                           &$\;\;\;\;\;$&
                  \multicolumn{6}{c}{{NL-SH:} {$K_{NM}=354.95$\,MeV}} \\
 &\,\,&\multicolumn{6}{c}{{SkM:} {$K_{NM}=216.7$\,MeV}}
                                                           &$\;\;\;\;\;$&
            \multicolumn{6}{c}{{\ \ \ \ SIII:} {$K_{NM}=356.00$\,MeV}} \cr
\cline{3-8} \cline{10-15}
Nuclei&\,\,&$K_C(A)$&$B(A)$&$B_{sr}/B_{rel}\;$&$\omega_C$&$\omega_{sr}$
&$\omega_{Sky}$
                                                           &$\;\;\;\;\;$
  &$K_C(A)$&$B(A)$&$B_{sr}/B_{rel}\;$&$\omega_C$&$\omega_{sr}$&$\omega_{Sky}$
\\
\cline{1-15}
        $^{16}$O&\,\,&\,\,74.0&0.2282&0.742&18.0&20.9&22.4&
                                     &105.6&0.2911&0.544&19.0&25.8&26.6 \\
          $^{40}Ca$&\,\,&102.0&0.3086&0.894&18.2&19.2&20.2&
                                     &153.4&0.3578&0.750&20.7&23.9&24.7 \\
          $^{90}Zr$&\,\,&117.1&0.4776&0.924&16.2&16.3&17.0&
                                     &149.0&0.4677&0.930&20.4&18.5&21.2 \\
         $^{208}Pb$&\,\,&116.0&0.7825&0.991&12.2&12.2&12.9&
                                     &197.7&0.8084&0.939&15.6&16.1&16.2 \\
\end{tabular}
\end{table}
%% END (TABLE ONE)%%%%%%%%%%%%%%%%%%%%%%%%%%%%%%%%%%%%%%%%%%
\hsize=15.5truecm

\figure{(a) The constrained incompressibility $K_C$ in Eq. (10) for a few
nuclei obtained using various RMF parameter sets. (b) The collective
mass $B_{rel}$ for the breathing-mode monopole vibrations obtained
from Eq. (16) within the RMF theory.}

\figure{The frequency $\omega_C$ of the monopole mode obtained
using eq. (17).}

\figure{The incompressibility $K_A$ (Eq. 20) obtained from 'scaling'
of the nuclear density in the semi-infinite nuclear matter using the
Thomas-Fermi approximation.}

\end{document}